# Thermodynamics: A Clear Conceptual Framework

Joaquim Anacleto

Departamento de Física, Escola de Ciências e Tecnologia, Universidade de Trás-os-Montes e Alto Douro, Quinta de Prados, 5000-801 Vila Real, Portugal

IFIMUP-IN e Departamento de Física e Astronomia, Faculdade de Ciências, Universidade do Porto, R. do Campo Alegre s/n, 4169-007 Porto, Portugal

e-mail: anacleto@utad.pt

ORCID: orcid.org/0000-0002-0299-0146

**Abstract**

Building on the fundamental equation, this study revisits key thermodynamic concepts in a cohesive and innovative manner. It demonstrates the consistency of thermodynamic theory while addressing and clarifying common misconceptions and errors found in the literature, particularly regarding discussions on heat and work. Although the latter two concepts could potentially be set aside, they can be retained if their various and different definitions are clearly articulated and properly understood. The proposed theoretical framework was tested using the free expansion of an ideal gas, a particularly demanding example due to its abrupt nature. From an educational standpoint, this article is invaluable as it consolidates fundamental yet often subtle concepts in an assertive and comprehensible way. Furthermore, it promotes a clearer and more accessible understanding of thermodynamics, challenging the widespread notion that is inherently difficult to grasp.

## 1. Introduction

Since I began delving into the theory of thermodynamics, two interesting quotes have stayed with me. One—which inspires and motivates me—comes from the physicist Albert Einstein:

*A theory is the more impressive the greater the simplicity of its premises, the more different kinds of things it relates, and the more extended its area of applicability. Therefore, the deep impression that classical thermodynamics made upon me. It is the only physical theory of universal content which I am convinced will never be overthrown, within the framework of applicability of its basic concepts.*

The other quote—by physicist Arnold Sommerfeld—highlights the nuanced complexity often found in thermodynamics:

*Thermodynamics is a funny subject. The first time you go through it, you don't understand it at all. The second time you go through it, you think you understand it, except for one or two small points. The third time you go through it, you know you don't understand it, but by that time you are so used to it, it doesn't bother you any more.*

Over time, I have found myself increasingly aligning with Einstein's perspective, while gradually moving away from Sommerfeld's view. Although the latter offers some comfort when grappling with the difficulties of thermodynamics, it became difficult for me to accept that debates persist about basic concepts in such a well-established branch of physics, often only to conclude that the concepts are indeed subtle.

Surprisingly, the concept of *heat*—which gives thermodynamics its distinct identity—is also the most prone to controversy and misunderstanding, frequently sparking intense debate [1–9]. This concept has undergone significant scientific evolution, shifting from the flawed caloric theory to the understanding of heat as energy in transit [10, 11]. However, is there already a scientific consensus on this concept? It is entirely acceptable for a concept to have varied interpretations, if they are all consistent with its definition, since interpreting often involves elements of subjectivity, mental visualisation, and intuition. However, from a strictly scientific perspective, the concept must have a clear and unique definition.

Beyond the concept of heat, other essential concepts such as reservoirs, work, dissipative work, lost work, exergy, the Clausius relation, entropy generation, and entropy flow—among others—also merit deeper exploration.

Let us liken thermodynamics to a high mountain. The evolution of its concepts and definitions corresponds to climbing the mountain, unexplored at first, following unknown paths, sometimes stepping back due to the appearance of difficult obstacles, then choosing alternative paths, eventually stopping to rest, resuming the climb with renewed enthusiasm, until finally reaching the summit. Once there, we can take in the setting and gain a comprehensive view of the mountain, allowing us to identify the best paths for descent, which are likely to be different from those used on the climb. This knowledge can then be applied to future climbs, enriching our understanding and approach.

The top of the mountain represents the fundamental equation of thermodynamics. With this vantage point, we can now descend the mountain while avoiding the mishaps experienced during the ascent. In other words, the fundamental equation—irrespective of how it was derived—serves as a tool to thoroughly explore and discuss various thermodynamic concepts.

The intuition that might have been used to choose the unknown paths that led us to the top is no longer necessary. The definitions and concepts based on the fundamental equation are obtained using only logical-mathematical reasoning, giving the results an undeniable consistency and robustness. This is what the remaining part of the article is dedicated to exploring.

## 2. Fundamental equation and process invariants

Thermodynamics is about macroscopic systems, establishing relations between their state properties. For a hydrostatic, single-component, one-phase and closed system, the internal energy $U = U(S,V)$ is a function of its natural variables entropy $S$ and volume $V$ [10–14]. The differential of this function is the *fundamental equation* in the energy representation,

$$dU = TdS - PdV, \qquad (1)$$

where $T = (\partial U/\partial S)_V$ is the temperature and $P = -(\partial U/\partial V)_S$ is the pressure.

The reason why (1) is regarded as a fundamental equation is that it encapsulates all essential thermodynamic information about the system [12]. A thermodynamic process is an interaction between the system and its surroundings, which is considered here as a sequence of infinitesimal steps to which (1) is applicable.

The physical laws remain unaffected regardless of whether we label a part of the thermodynamic universe as the 'system' or 'surroundings'. Thus, (1) can also be applied to describe the surroundings and distinguish its properties and variables from those of the system, whereby the subscript 'e' (denoting external) is used. The lack of clear notation to make this distinction is common in the literature and often leads to confusion when interpreting thermodynamic concepts. Therefore, applying (1) to the surroundings, we obtain

$$dU_e = T_e dS_e - P_e dV_e. \qquad (2)$$

Conservation of energy during a process is stated as

$$dU + dU_e = 0, \qquad (3)$$

which—together with (1) and (2)—establishes the *fundamental equation for the process*

$$TdS - PdV = -T_e dS_e + P_e dV_e. \qquad (4)$$

Resulting from the combination of conservation of energy with the fundamental equations for the system and surroundings, (4) contains all essential thermodynamic information about the process. All definitions and concepts should be consistent with this equation.

Entropy is widely regarded as one of the most challenging concepts in thermodynamics, not because it is more abstract than energy but perhaps because—unlike energy—it is generally not conserved. Indeed, the *second law of thermodynamics* states that while the total entropy remains constant in reversible processes, it increases in irreversible ones. Thus, there is no counterpart to (3) for entropy; rather, we have

$$dS + dS_e = dS_G \geq 0, \qquad (5)$$

in which $dS_G$ stands for the *entropy generation* in the infinitesimal process described by (4).

From (1)–(3), $dS$ and $dS_e$ can be written as

$$dS = \frac{dU}{T} + \frac{PdV}{T}, \qquad (6)$$

$$dS_e = -\frac{dU}{T_e} + \frac{P_e dV_e}{T_e}, \qquad (7)$$

such that the entropy generation $dS_G$ can be expressed as

$$dS_G = \left(\frac{1}{T} - \frac{1}{T_e}\right)dU + \frac{PdV}{T} + \frac{P_e dV_e}{T_e} \geq 0. \qquad (8)$$

Equations (3), (4), (5), and (8) are *process invariants* [15, 16], which means that they have the notable property that system and surroundings variables are interchangeable. When the index 'e' is moved from surroundings to system variables, the equations remain unchanged, thus emphasising its essential nature. Dividing the thermodynamic universe into two interacting parts, these equations apply regardless of which part is chosen as the system. They are symmetric with respect to the interchange between the system and its surroundings.

At this juncture, it becomes pertinent to question the role of the concepts of *heat* and *work*. Their absence thus far suggests either a lack of necessity or that their definitions fail to emerge from the preceding equations. We could critically wonder whether these concepts are indeed indispensable. Given the issues that they cause, it might be worth considering their removal from the theory of thermodynamics [3, 5, 6, 8].

However, the formulation of the *first law of thermodynamics* is centred on the concepts of heat $Q$ and work $W$. In its differential form, this law is expressed as

$$dU = \delta Q + \delta W. \qquad (9)$$

By inserting (9) into (8), we obtain the entropy generation as

$$dS_G = \left(\frac{1}{T} - \frac{1}{T_e}\right)\delta Q + \frac{\delta W + PdV}{T} + \frac{P_e dV_e - \delta W}{T_e} \geq 0. \qquad (10)$$

Although heat $\delta Q$ and work $\delta W$ have been forced to appear in (10) through the first law, they do not introduce any additional information into (8). The validity of (10) holds regardless of how heat and work are defined, or even if they are removed from theory. Heat and work are not fundamental constructs in thermodynamics as they do not derive directly from first principles but instead reflect intuitive, context-dependent interpretations. In the absence of additional assumptions, their definitions remain indeterminate. This issue will be addressed in the sections that follow.

## 3. Heat, work, and reservoirs

Interestingly, although heat is a cornerstone of thermodynamic theory, it is precisely this concept that often gives rise to the most significant challenges and misunderstandings. The notion of heat has a deep-rooted familiarity, shaped by everyday experiences and discussed over many years of thought and observation. The intuitive nature of this concept finds scientific validity only in reversible processes. For all others—which are the vast majority— we must abandon our intuition to construct a coherent and unambiguous theoretical framework.

For a given process, it would be desirable to know *where entropy is generated* and obtain the *entropy flow* between the system and its surroundings. The entropy flow depends on where entropy generation takes place, although this location is intrinsically indeterminate. Regarding entropy, only $dS$ and $dS_e$ are defined.

Unlike an invariant, a *flow* changes sign when a system-surroundings interchange occurs. Thus, combining (6), (7), and (9), and introducing the *entropy flow* $dS_\phi$, $dS$ and $dS_e$ become

$$dS = \frac{\delta Q}{T} + \frac{\delta W + PdV}{T} = dS_\phi + \beta\, dS_G, \tag{11}$$

$$dS_e = -\frac{\delta Q}{T_e} + \frac{P_e dV_e - \delta W}{T_e} = -dS_\phi + \beta_e\, dS_G, \tag{12}$$

where $\beta$ and $\beta_e$ are the entropy generation fractions within the system and its surroundings, respectively, thus satisfying

$$\beta, \beta_e \in [0,1] \text{ and } \beta + \beta_e = 1. \tag{13}$$

It is worth emphasising that the *entropy flow* does not contribute to the *entropy generation*. Additionally, $dS_\phi$ depends on the values of $\beta$ and $\beta_e$, which can be freely selected provided that they comply (13). Rather than representing a deficiency in the theory, this indeterminacy highlights that the location of entropy generation has no inherent physical meaning but is established by convention.

Throughout the development of thermodynamics, it has often been assumed that entropy generation occurs exclusively within the system. This assumption is equivalent to selecting

$$\beta = 1 \text{ and } \beta_e = 0, \tag{14}$$

which corresponds to considering surroundings constituted by *reservoirs* [17]. Thus, (11) and (12) become

$$dS = \frac{\delta Q}{T} + \frac{\delta W + PdV}{T} = dS_\phi + dS_G, \tag{15}$$

$$dS_e = -\frac{\delta Q}{T_e} + \frac{P_e dV_e - \delta W}{T_e} = -dS_\phi. \tag{16}$$

Subsequently, a natural definition of $\delta Q$ is to make it proportional to $dS_\phi$, which—according to (16)—implies that work and heat should be defined as

$$\delta W = P_e dV_e, \tag{17}$$

$$\delta Q = T_e dS_\phi = -T_e dS_e. \tag{18}$$

Therefore, with these work and heat definitions, from (15) and (16), $dS$ and $dS_e$ can be written as

$$dS = \frac{\delta Q}{T} + \frac{\delta W_D}{T}, \tag{19}$$

$$dS_e = -\frac{\delta Q}{T_e}, \tag{20}$$

where $\delta W_D$ is the *dissipative work*, as a *process invariant* given by

$$\delta W_D = PdV + P_e dV_e. \tag{21}$$

The definitions (17) and (18) for $\delta W$ and $\delta Q$ are not the only possible ones but they are definitely the only ones consistent with the *Clausius relation* [10, 18]. This relation can be derived by integrating (5) over a cyclical process—one in which the system returns to its initial state. Thus, we have

$$\oint (dS + dS_e) = \oint dS_e \geq 0, \tag{22}$$

which—combined with (18)—immediately gives the Clausius relation

$$-\oint dS_e = \oint \frac{\delta Q}{T_e} \leq 0. \tag{23}$$

The Clausius relation (23) is considered by many as the most important relation in classical physics. Here, it has been obtained elegantly by combining the formalism resulting from the fundamental equation with the definitions (17) and (18) for work and heat. The important point to keep in mind is that with definitions other than those that have been chosen, the relation (23) does not hold.

**4. Lost work and exergy**

Inserting (17) into (10), multiplying by it $T$, and using (21), we obtain the *lost work* $TdS_G$ as

$$TdS_G = \left(1 - \frac{T}{T_e}\right)\delta Q + \delta W_D, \tag{24}$$

where the quantity that multiplies $\delta Q$ is the *Carnot efficiency* $\eta_C$, given by

$$\eta_C = 1 - \frac{T}{T_e}. \tag{25}$$

The first term on the right-hand side of (24) corresponds to *thermal irreversibility*, while the second term—which equals dissipative work—concerns *mechanical irreversibility* [16]. Curiously, the *Carnot efficiency* $\eta_C$—typically introduced in the context of heat engines—is here derived more generally and linked to the lost work in an irreversible process.

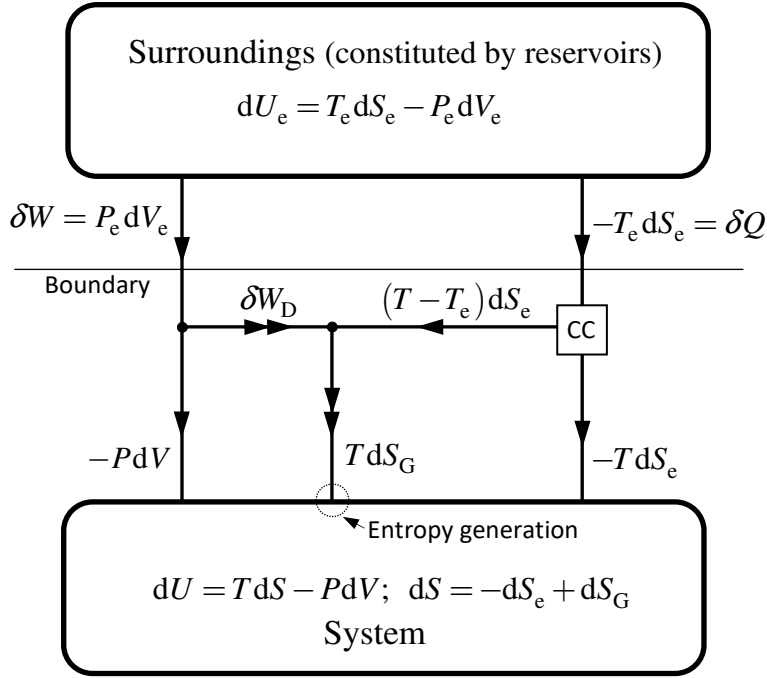

**Figure 1.** Conceptual structure for a general thermodynamic process when irreversibility is assumed to occur within the system. CC = Carnot cycle. The seven lines represent energies in transit, whose conceptual meanings are given by the expressions next to them (see the text for more details). Table 1 assists this figure by highlighting certain concepts.

**Table 1.** Relevant concepts contained in Figure 1. Irreversibility in the system and thus its surroundings are constituted by reservoirs.

| | |
|---|---|
| $P_e dV_e = \delta W$ | work |
| $-T_e dS_e = \delta Q$ | heat |
| $-dS_e = dS_\phi$ | entropy flow |
| $dS_G = \left(\dfrac{1}{T} - \dfrac{1}{T_e}\right)\delta Q + \dfrac{\delta W_D}{T}$ | entropy generation (invariant) |
| $\delta W_D = PdV + P_e dV_e$ | dissipative work (invariant) |
| $(T - T_e)dS_e = \left(1 - \dfrac{T}{T_e}\right)\delta Q$ | work delivered or absorbed by CC |
| $TdS_G = (T - T_e)dS_e + \delta W_D$ | lost work |

The reason for calling $T\mathrm{d}S_\mathrm{G}$ *lost work* is that it refers to the additional work that could have been achieved if the process had been made reversible through proper modifications to the system alone.

Figure 1 shows the conceptual structure for a general thermodynamic process when irreversibility is present within the system and thus the surroundings are constituted by reservoirs. Each of the seven lines represents energy in transit, defined by the adjacent expression. For the single-arrowed lines, energy values are positive when travelling in the direction of the arrows and negative when travelling opposite to them. In contrast, for the double-arrowed lines, energy values are always positive and travel in the direction of the arrows. Naturally, some of these energy values might be zero, depending on the process. Table 1 complements Figure 1 by summarising these concepts.

As previously discussed, the question of where irreversibility occurs is irrelevant within the theory. Therefore, we can think of it wherever we want. Considering it in the system—as we did before—through the assignments (14), we obtain the definitions of work and heat as (17) and (18), which imply the conceptual structure shown in Figure 1. However, if we choose irreversibility in the surroundings instead of the system, which is equivalent to taking

$$\beta = 0 \text{ and } \beta_\mathrm{e} = 1, \tag{26}$$

reasoning parallel to the above leads us to the following different definitions of work and heat

$$\delta W' = -P\mathrm{d}V, \tag{27}$$

$$\delta Q' = T\mathrm{d}S. \tag{28}$$

Inserting these new definitions in (10), multiplying it by $T_\mathrm{e}$, and using (21), we obtain the lost work $T_\mathrm{e}\mathrm{d}S_\mathrm{G}$, which is known in literature as *exergy* [19-21]

$$T_\mathrm{e}\mathrm{d}S_\mathrm{G} = \left(\frac{T_\mathrm{e}}{T} - 1\right)\delta Q' + \delta W_\mathrm{D}. \tag{29}$$

This perspective is adopted especially in engineering [16, 19-21], a field in which thermodynamics has many applications. By using (21) and assuming that $\mathrm{d}V_\mathrm{e} = -\mathrm{d}V$, $T_\mathrm{e} = T_0 = \text{constant}$, and $P_\mathrm{e} = P_0 = \text{constant}$, (29) can be rewritten as

$$T_0\mathrm{d}S_\mathrm{G} = \left(\frac{T_0}{T} - 1\right)T\mathrm{d}S + P\mathrm{d}V - P_0\mathrm{d}V. \tag{30}$$

If the system evolves until it reaches the thermodynamic equilibrium with the surroundings, integrating (30) yields the *exergy* as it is typically presented in engineering

$$\int T_0\mathrm{d}S_\mathrm{G} = (U - U_0) - T_0(S - S_0) + P_0(V - V_0), \tag{31}$$

where $U_0$, $S_0$, and $V_0$ are values for the system upon reaching equilibrium.

Figure 1 requires adjustments to depict the conceptual structure that results from the new definitions (27) and (28) for work and heat. Specifically, the line that represents the lost work needs to be redirected to the surroundings, and it is necessary to update the expressions for certain energies. These changes result in Figure 2 and the supporting Table 2.

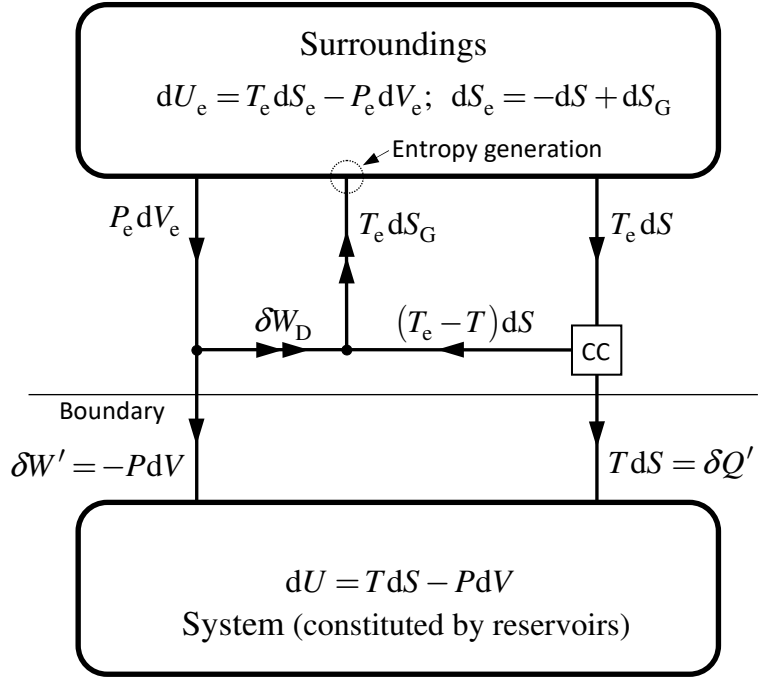

**Figure 2.** Conceptual structure for a general thermodynamic process when irreversibility is assumed to occur within the surroundings. CC = Carnot cycle. The seven lines represent energies in transit, whose conceptual meanings are given by the expressions next to them (see the text for more details). Table 2 assists this figure by highlighting certain concepts.

**Table 2.** Relevant concepts contained in Figure 2. Irreversibility in the surroundings and thus the system is constituted by reservoirs.

| | |
|---|---|
| $-PdV = \delta W'$ | work |
| $TdS = \delta Q'$ | heat |
| $dS = dS_\phi$ | entropy flow |
| $dS_G = \left(\dfrac{1}{T} - \dfrac{1}{T_e}\right)\delta Q' + \dfrac{\delta W_D}{T_e}$ | entropy generation (invariant) |
| $\delta W_D = PdV + P_e dV_e$ | dissipative work (invariant) |
| $(T_e - T)dS = \left(\dfrac{T_e}{T} - 1\right)\delta Q'$ | work delivered or absorbed by CC |
| $T_e dS_G = (T_e - T)dS + \delta W_D$ | lost work (known as exergy) |

As the changes $dU$, $dS$, $dU_e$, and $dS_e$ (and therefore $dS_G$) are the same in Figures 1 and 2, they in fact represent the same thermodynamic process [22]. However, the values of *heat*, *work*, *lost work*, and *entropy flow* differ in the two cases because they depend on where entropy generation is assumed to occur, within the system or surroundings. These quantities are not state functions and have no meaning once the process is complete.

It is also important to clarify that the Clausius relation is not consistent with definitions (27) and (28) because (23) does not hold when $\delta Q$ is replaced by $\delta Q'$. Indeed, by combining (4), (21) and (28), $\delta Q'/T_e$ can be written as

$$\frac{\delta Q'}{T_e} = \frac{\delta W_D}{T_e} - dS_e, \tag{32}$$

which—upon integration for a cyclical process—gives

$$\oint \frac{\delta Q'}{T_e} = \oint \frac{\delta W_D}{T_e} + \oint (-dS_e). \tag{33}$$

Finally, using (22), we obtain

$$\oint \frac{\delta Q'}{T_e} \leq \oint \frac{\delta W_D}{T_e}, \tag{34}$$

which is definitively not the Clausius relation.

Although the conceptual discussion could conclude here without compromising the paper's completeness, the following section is included to deepen understanding through a concrete and challenging example before proceeding to the conclusions.

## 5. Illustrative example: Free expansion of an ideal gas

This final section explores the classic example of *free expansion of an ideal gas* [10, 11] through the lens of the conceptual framework developed earlier. Despite its apparent simplicity, this process presents significant conceptual challenges and serves as a stringent test of thermodynamic reasoning. Being irreversible and non-quasistatic, free expansion raises important questions regarding the proper definition and interpretation of quantities such as heat, entropy generation, and lost work [23].

An initial and critical challenge concerns how to meaningfully define and handle thermodynamic variables and their differentials during such an inherently violent process. To address this difficulty, the approach detailed in [24] is used, whereby the standard free expansion is replaced by a sequence of infinitesimal free expansions, carried out at a sufficiently slow rate to yield a quasistatic process while preserving all thermodynamic properties of the original process, namely, the same initial and final states and entropy generation. From a thermodynamic standpoint, the standard free expansion (violent in nature) and the sequence of infinitesimal free expansions (quasistatic) are indistinguishable processes [22]. Nonetheless, the latter holds the crucial advantage of featuring well-defined variables and their differentials throughout the entire process. The replacement of a non-quasistatic process by an identical quasistatic one is a procedure that can in principle be extended to all such processes.

Free expansion can therefore be conceptually modelled as a quasistatic process comprising infinitesimal volume increments $dV > 0$, each resulting from the sequential removal of an arbitrarily large number of equally spaced barriers within an initially evacuated volume [24]. Within the conceptual framework proposed above, even a seemingly peculiar process such as the free expansion of an ideal gas can be coherently interpreted through its quasistatic proxy. Figure 3 results from the application of Figure 1 to this process.

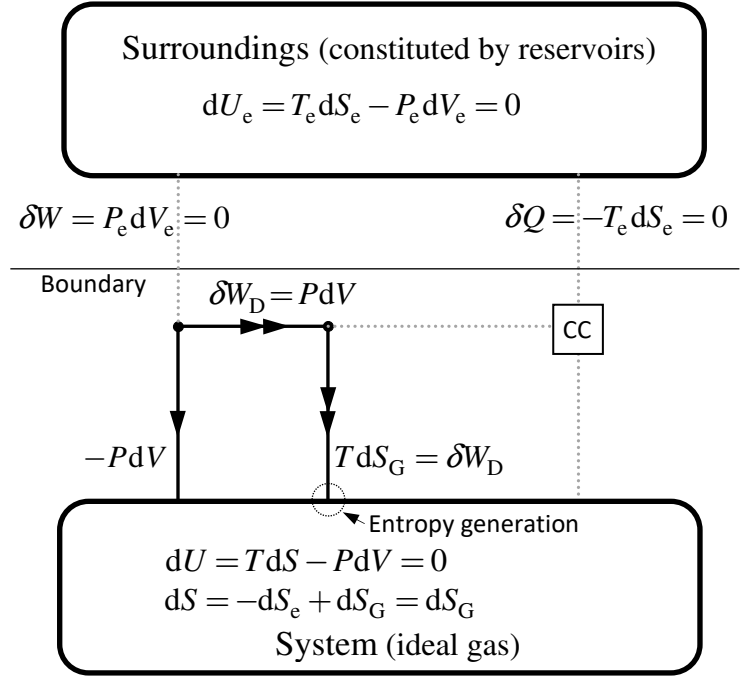

**Figure 3.** Conceptual structure for the free expansion of an ideal gas when irreversibility is assumed to occur within the system.

Although one might argue that there is no interaction with the surroundings, it is more compelling to adopt a broader conceptual view and regard this process as a specific type of interaction in which nothing occurs in the surroundings, analogous to the role that zero plays in mathematics. Therefore, since there is no exchange with the surroundings during this process, we have $\delta W = 0$, $\delta Q = 0$, $dU = 0$, $dU_e = 0$, and $dS_e = 0$. By (18), $dS_\phi = -dS_e$. This implies that there is no entropy flow but only entropy generation, which therefore matches the change in system entropy, namely $dS_G = dS$. The lost work $T dS_G$ is simply expressed as

$$T_e dS_G = \delta W_D = P dV = T dS .\tag{35}$$

Since $dU = 0$ and internal energy depends only on temperature, it remains constant during the free expansion. Moreover, for an ideal gas, the state equation is given by

$$PV = nRT ,\tag{36}$$

where $n$ is the number of moles and $R$ is the universal gas constant.

Accordingly, from (35) and (36), we obtain

$$dS = nR \frac{dV}{V}, \qquad (37)$$

which—upon integration over a volume change from $V_1$ to $V_2$—yields a well-known result:

$$\Delta S = nR \ln \frac{V_2}{V_1}. \qquad (38)$$

Could the irreversibility—or, equivalently, the entropy generation—be attributed instead to the surroundings, despite no change occurring in there? This is conceptually possible if the changes in the surroundings cancel out, keeping the entropy generation the same. Accordingly, we apply the conceptual framework of Figure 2, which leads to the interaction depicted in Figure 4, where the irreversibility unambiguously takes place in the surroundings.

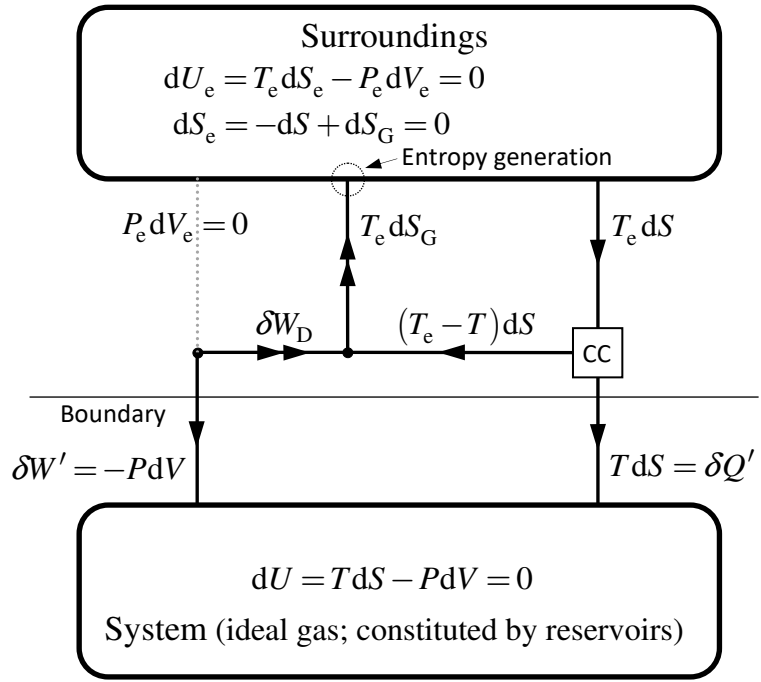

**Figure 4.** Conceptual structure for the free expansion of an ideal gas when irreversibility is assumed to occur within the surroundings.

With (27) and (28) now applicable, we have $\delta W' = -PdV$ and $\delta Q' = TdS$. Moreover, the lost work $T_e dS_G$ becomes

$$T_e dS_G = PdV + (T_e - T)dS. \qquad (39)$$

Since $dU = TdS - PdV = 0$ still holds, (39) reduces to

$$T_e dS_G = T_e dS, \qquad (40)$$

and since $dS_\phi = dS$ (see Table 2), we obtain

$$dS_G = dS = dS_\phi. \qquad (41)$$

Evidently, (37) and (38) remain valid. Furthermore, it is worth noting that all expressions are valid regardless of the surroundings temperature $T_e$, namely whether $T_e > T$, $T_e < T$, or $T_e = T$.

The entropy generation $dS_G$ is the same—as it must—although in this description, it occurs in the surroundings, namely there is no irreversibility within the system. What differs now is the presence of heat $\delta Q'$, work $\delta W'$, and entropy flow $dS_\phi$ across the boundary, and the fact that the lost work is given by $T_e dS_G$ instead of $T dS_G$. These differences do not pose any problem as the four quantities in question cease to have any meaning once the process is complete. Thus, the two descriptions—Figures 3 and 4—are conceptually indistinguishable.

In contrast to Figure 3, it is understandable that Figure 4 might be met with some resistance as a valid representation of the free expansion of an ideal gas. Nevertheless, both descriptions yield the same final state of the system, leave the surroundings unaltered, and result in the same entropy generation. Upon completion of the process, no observable criterion allows us to determine which description was actually used.

## 6. Conclusions

This work has presented a clear, coherent, and integrative theoretical framework encompassing the principles, concepts, and definitions of thermodynamics. From this foundation, a general and non-specific process—either reversible or irreversible—has been illustrated through Figures 1 and 2, offering significant instructional value and allowing the conceptual structure and interrelationships among key ideas to be clearly and comprehensively understood. Within this framework, the free expansion of an ideal gas was examined not as a typical case but a peculiar and demanding example, chosen specifically to test the coherence and robustness of the proposed conceptual structure.

We believe that this work contributes not only to clarifying persistent misconceptions and errors in the literature but also to challenging the common perception that thermodynamics is inherently difficult. At the very least, we hope that it encourages more critical and reflective engagement with the subtle conceptual issues that are central to thermodynamics.

**Data availability statement**

No new data were created or analysed in this study.